\documentclass[twocolumn]{aastex61}
\received{July 1, 2016}
\revised{September 27, 2016}
\accepted{\today}
\submitjournal{ApJ}
\shorttitle{Supercritical Accretion onto a Neutron Star}
\shortauthors{H. R. Takahashi, S. Mineshige \& K. Ohsuga}

\begin{document}

\title{Supercritical Accretion onto a Non-Magnetized Neutron Star: Why is it Feasible?}
\correspondingauthor{Hiroyuki Takahashi}
\email{takahashi@cfca.jp}

\author[0000-0003-0114-5378]{Hiroyuki R. Takahashi}
\affil{Center for Computational Astrophysics, National
  Astronomical Observatory of Japan, National Institutes of Natural
  Sciences, Mitaka, Tokyo 181-8588, Japan}

\author{Shin Mineshige}
\affil{Department of Astronomy, Kyoto University, Oiwake-cho, Kitashirakawa, Sakyo-ku, 
Kyoto 606-8502, Japan}

\author{Ken Ohsuga}
\affil{Division of Theoretical Astronomy, National
  Astronomical Observatory of Japan,
  National Institutes of Natural
  Sciences, Mitaka, Tokyo 181-8588, Japan}
  \affil{School of Physical Sciences,Graduate University of
Advanced Study (SOKENDAI), Shonan Village, Hayama, Kanagawa 240-0193,
Japan}

\begin{abstract}
 To understand why supercritical accretion is feasible onto a neutron
 star, we carefully examine the accretion flow dynamics by
 2.5-dimensional general relativistic (GR) radiation magnetohydrodynamic
 (RMHD) simulations, comparing the cases of accretion onto a
 non-magnetized neutron star (NS) and that onto a black hole
 (BH). Supercritical BH accretion is relatively easy, since BH can
 swallow excess radiation energy, so that radiation flux can be inward
 in its vicinity. This mechanism can never work for NS which has a solid
 surface. In fact, we find that the radiation force is always
 outward. Instead, we found significant reduction in the mass accretion
 rate due to strong radiation-pressure driven outflow. 

The radiation flux $F_\mathrm{rad}$ is self-regulated such that
 the radiation force balances with the sum of gravity and centrifugal
 forces. Even when the radiation energy density much exceeds that
 expected from the Eddington luminosity 
$E_\mathrm{rad}
 \simeq F_\mathrm{rad}\tau/c> 10^2 L_\mathrm{Edd}/(4\pi r^2 c)$,
the radiation flux is always kept below the certain value 
which makes it possible not to blow all the gas away from the disk. 
 These effects make supercritical accretion feasible.
 We also find that a settling region,
 where accretion is significantly decelerated by radiation cushion, is
 formed around the NS surface. 
In the settling region, the radiation
 temperature and mass density roughly follow $T_\mathrm{rad} \propto
 r^{-1}$ and $\rho \propto r^{-3}$, respectively.
 No settling region appears
 around the BH so that matter can be directly swallowed by the BH with
 supersonic speed. 
\end{abstract}




\section{Introduction} \label{sec:intro}
There are growing evidences recently of the supercritical (or super-Eddington) accretion objects
(hereafter, super-Eddington accretors) in the Universe. Super-Eddington
accretors are very powerful engines and so play essential roles in
various astrophysical phenomena (e.g., emitting high energy emission
and/or launching relativistic baryon jets). They can also give large
impacts on their environments through intense radiation and massive
outflow, thereby giving rise to interesting activities (e.g., creating
huge ionized nebulae). It is thus worth of studying the detailed
processes associated with super-Eddington accretors from various
viewpoints.

One of the most promising candidates for the super-Eddington accretors
is ULXs, compact Ultraluminous X-ray sources, which were successively
discovered in nearby active galaxies 
\citep{1989ApJ...347..127F,2011ApJS..192...10L,2011MNRAS.416.1844W}. 
The ULXs are off-nuclear point sources producing
very large X-ray luminosity, $L_{\rm x} > 10^{39}$ erg s$^{-1}$, far
exceeding the Eddington limit ($L_{\rm Edd}$) of a stellar mass black
hole. 
There are two major
scenarios so far proposed and discussed to explain their nature: (1)
sub-Eddington accretors harboring an intermediate mass black hole (IMBH)
with mass exceeding $100 M_\odot$ \citep[e.g.][]{2000ApJ...535..632M,
2004ApJ...614L.117M}, and (2) super-Eddington accretors harboring a
stellar mass black hole with super-Eddington rates with ${\dot M} \gg L_{\rm Edd}/c^2$
\citep[e.g.][]{2001ApJ...549L..77W,2001ApJ...552L.109K,2007MNRAS.377.1187P}. Quite
recently, one very
convincing piece of evidence in favor of the latter scenario has been
reported; that is the discovery of pulses in one of the ULXs M82 X-2
\citep{2014Natur.514..202B}. This discovery has established that at least
some part of ULXs is super-Eddington accretors 
\citep[ULX Pulsars, see][for the discovery of other 
cases]{2016ApJ...831L..14F,2017Sci...355..817I,2017MNRAS.466L..48I}.

The ULXs are not the only candidate for super-Eddington accretors,
however, there are actually plenty of other objects known to date, that
are suspected to host supercritical accretion flow. One good example is
ULSs, Ultraluminous supersoft sources, which have similarly high X-ray
luminosities but which exhibit much softer X-ray spectra with typical
photon energy of $\sim 0.1$ keV 
\citep[e.g.,][]{2003ApJ...592..884D,2004ApJ...617L..49K}. These features
can be understood, if one observes
super-Eddington accretors from nearly edge-on direction
\citep{2016MNRAS.456.1859U,2016ApJ...818L...4G,2017PASJ..tmp..143O}. 
Other candidates include microquasars, TDE (tidally disrupted events),
narrow-line Seyfert 1 galaxies
\citep{1999ApJ...522..839W,2000PASJ...52..499M}, and so
on. Super-Eddington accretors are unique in the sense that 
their energy release rate does not depend on their internal properties
at all but on the external conditions;
i.e., mass supply rate to the compact object vicinity.

In parallel with accumulation of observational evidences supporting the
ubiquitous existence of super-Eddington accretors, semi-analytic and
simulation studies have been conducted rather extensively in these days.
The possibility of supercritical accretion onto the compact star was
first discussed in the pioneering paper by \cite{1973A&A....24..337S}
(herefter SS73).
\cite{1988ApJ...332..646A} found an equilibrium solution
of the supercritical disk and constructed the so-called slim disk model,
in which advection of radiation entropy plays a crucial role 
\citep[see][for a simplified self-similar solution of the slim
disk]{1999PASJ...51..725W}.
The general relativistic version of the slim disk was first constructed
by \cite{1998MNRAS.297..739B}, who claimed that the thermalization
timescale could be longer than the accretion timescale so that radiation
and matter temperatures may deviate.
The supercritical accretion disk has also been discussed in the context
of magnetized and/or non-magnetized neutron star. In the case of
accretion onto a magnetized neutron star, the accretion mode through the
disks quenches due to the strong magnetic pressure. Gas then falls onto
the neutron star surface along the magnetic field lines, thereby forming
accretion columns \citep{1976MNRAS.175..395B,1988SvAL...14..390L}.
The emission from the accretion columns can reaches 
$10^{40}\ \mathrm{erg s^{-1}}$
\citep{2015MNRAS.454.2539M}, 
which is consistent with resent observation of the ULX pulsars.

The pioneering simulation work was made by \cite{2005ApJ...628..368O} using
radiation hydrodynamic (RHD) simulations.
They could for the first time succeed in producing steady-state supercritical accretion flow
and revealed various unique features, such as anisotropic radiation field, wide-angle outflow,
large-scale circulation of gas within the flow, and so on.
The most up-to-dated simulations are performed under the full GR treatments including magnetic field for BH
\citep{2014MNRAS.441.3177M,2014MNRAS.439..503S,2015MNRAS.454.2372S,2016MNRAS.456.3929S,2016ApJ...826..23,2017MNRAS.466..705S}
and for NS \citep{2017ApJ...845L...9T}, and found formation of strong
outflows \citep{2015PASJ...67...60T,2015MNRAS.453.3213S}.
\cite{2016ApJ...826..23} demonstrate that the hot accretion flow is
formed closed to the compact object and it can be responsible for hard
X-ray emission.

We, here, wish to address one key question; why is supercritical accretion feasible?
Another related question is; is there no practical limits on mass accretion rates and luminosities,
provided that sufficient amount of mass is supplied externally?
Through the number of simulation studies conducted recently we now have a consensus
that it is really feasible to put material into a BH as much as you like.
We should be careful, however, since the simulations only give results, 
while it is our task to specify mechanisms underlying them.
Popular argument made in this context is as follows:
supercritical accretion is feasible, since radiation goes out in the perpendicular direction to the disk plane, 
thus giving little effects on the matter that accretes along the disk plane.
This explanation is not complete, however, since 
it misses the consideration of the force balance on the equatorial plane,
although radiation force should also give enormous impacts on the material there.
What is needed is to give a clear explanation why matter can accrete towards the region full of radiation energy.

It is interesting to note in this respect that 
\cite{2007ApJ...670.1283O} discussed this problem, by using their RHD simulation data.
They have found two key ingredients which make it possible to excite supercritical flow:
anisotropic radiation field created by large $\tau$ accretion flow from the equatorial plane
and photon trapping effects; photons created deep inside the thick accretion flow are trapped within the
flow and finally swallowed by a BH before escaping from the surface of the flow.
The outgoing radiative flux is thus largely attenuated (or sometimes flux becomes inward)
so that supercritical accretion is feasible onto BHs.

How about the cases of NS accretions? 
We should point that photon trapping cannot be so effective on a long timescale there, 
since photons should eventually be emitted from the solid surface of a NS.
As a result, radiation force should always be outward, thereby decelerating accreting gas.
Supercritical accretion is relatively easier, if the NS is strongly magnetized
and if accretion occurs through a narrow accretion column (i.e., ULX pulsars). This is because
excess radiation energy can then almost freely escape from the side wall
of the accretion column 
\citep{1976MNRAS.175..395B,2016PASJ...68...83K,2017ApJ...845L...9T}.
In this paper, we make careful analysis of the GR simulation data
to find an answer to the question, why the super-Eddington accretion
onto a non-magnetized NS is feasible. The paper is organized as follows:
we will describe the methods of calculations in section 2 and then present results in section 3.
Final section is devoted to discussion on observational implications and other related issues.

\section{Basic Equations and Numerical Method}
We numerically solve general relativistic Radiation Magnetohydrodynamic (GR-RMHD) equations,
in which the radiation equation is based
on a moment formalism with applying a M-1 closure
\citep{1984JQSRT..31..149L, 2013MNRAS.429.3533S, 2013PASJ...65...72K}.
In the following, Greek suffixes indicate space-time components, and
Latin suffixes indicate space components. 
We take the light speed $c$ as unity otherwise stated. Then length and time are
normalized by gravitational radius $r_\mathrm{g}=GM/c^2$ and its light
crossing time $t_\mathrm{g}=r_\mathrm{g}/c$, where $G$ is the
gravitational constant and $M$ is a mass of a central object.
We take $M=1.4M_\odot$ and $M=10 M_\odot$ for NS and BH, respectively.

Basic equations consist of mass conservation,
\begin{equation}
 (\rho u^\nu)_{;\nu} = 0,\label{eq:masscons}
\end{equation}
the energy momentum conservation for magnetofluids,
\begin{equation}
 T^\nu_{\mu;\nu}= G_\mu, \label{eq:mhdeq}
\end{equation}
the energy momentum tensor for radiation field,
\begin{equation}
 R^\nu_{\mu;\nu} = -G_\mu,
  \label{eq:radeq}
\end{equation}
and induction equation,
\begin{equation}
 \partial_t (\sqrt{-g}B^i)=[\sqrt{-g}(B^i v^j - B^j v^i)],
  \label{eq:induction}
\end{equation}
where $\rho$ is the proper mass density, $u^\nu$ is the four velocity,
$v^i=u^i/u^0$ is the laboratory frame three velocity, $B^i$ is the
laboratory frame magnetic three field, and $g=\det({g_{\mu\nu}})$ is the
determinant of metric, $g_{\mu\nu}$.

The energy momentum tensor for magnetofluid and radiation are given by
\begin{eqnarray}
T^{\mu \nu}&=&
 \left(\rho + e + p_\mathrm{gas} + 2 p_\mathrm{mag}\right)u^\mu u^\nu
\nonumber \\
 &&+\left(p_\mathrm{gas} +p_\mathrm{mag}\right)g_{\mu\nu}
 -b^\mu b^\nu,\\
R^{\mu\nu}&=& p_{\mathrm{rad}}
\left(4u^\mu_\mathrm{rad} u^\nu_\mathrm{rad} + g^{\mu\nu}
 \right),
\end{eqnarray}
where $p_\mathrm{gas}, e, p_\mathrm{mag}, p_\mathrm{rad}$
and $u_\mathrm{rad}^\mu$ are the gas pressure,
gas internal energy, magnetic pressure,
radiation pressure,
and radiation frame's four velocity.
The gas internal energy is related to the gas pressure by
$e=(\Gamma-1)p_\mathrm{gas}$ with $\Gamma=5/3$ being the specific heat ratio.
The magnetic four vector $b^\mu$ is related to its three
vector through $b^\mu = B^\nu h_\nu^\mu/u^0$, where
$h_\nu^\mu=\delta_\nu^\mu + u_\mu u^\nu$ is the projection tensor and
$\delta^\mu_\nu$ is the Kronecker delta. The magnetic pressure is
represented by $p_\mathrm{mag}=b_\mu b^\mu/2$.

The gas and radiation field interact each other through a radiation
four force $G^\mu$, which is represented by
\begin{eqnarray}
 G^\mu&=& -\rho\kappa_\mathrm{abs}(R^\mu_\alpha u^\alpha + 4 \pi
  \mathrm{B}u^\mu)\nonumber \\
 &-& \rho\kappa_\mathrm{sca}
  (R^\mu_\alpha u^\alpha + R^\alpha_\beta u_\alpha u^\beta u^\mu)
  +G^\mu_\mathrm{comp},
\end{eqnarray}
where $\kappa_\mathrm{abs}=6.4\times 10^{22}\rho T_\mathrm{gas}^{-3.5}\
\mathrm{cm\ g^{-1}}$
and $\kappa_\mathrm{sca}=0.4\ \mathrm{cm\ g^{-1}}$ are free-free absorption and
Thomson-scattering opacities.
The gas temperature is calculated by 
$T_\mathrm{gas}=\mu m_p p_\mathrm{gas}/\rho k_\mathrm{B}$,
where $m_p$ is the proton mass,
$k_\mathrm{B}$ is the Boltzmann constant,
and $\mu=0.5$ is the mean molecular weight. 
The black body intensity is given by
$\mathrm{B}=a_\mathrm{rad} T^4_\mathrm{gas}$ with
$a_\mathrm{rad}$ being the radiation constant. 
We included the thermal Comptonization as follows:
\begin{eqnarray}
G^\mu_\mathrm{comp}
 &=& -\rho\kappa_\mathrm{sca} \hat{E}_\mathrm{rad}
 \frac{4k_\mathrm{B}(T_\mathrm{e}-T_\mathrm{rad})}{m_\mathrm{e}}
 \nonumber \\
&\times& \left[
  1 + 3.683\left(\frac{k_\mathrm{B} T_\mathrm{e}}{m_\mathrm{e}}\right)
  + 4 \left(\frac{k_\mathrm{B} T_\mathrm{e}}{m_\mathrm{e}}\right)^2
 \right]\nonumber \\
&\times& \left[
  1 + \left(\frac{k_\mathrm{B} T_\mathrm{e}}{m_\mathrm{e}}\right)
 \right]^{-1}u^\mu,
\end{eqnarray}
where $T_\mathrm{e}$ is the electron temperature, $\hat{E}_\mathrm{rad}$
is the comoving frame radiation energy density,
$T_\mathrm{rad}=(\hat{E}_\mathrm{rad}/a_\mathrm{rad})^{1/4}$ is the
radiation temperature, and $m_\mathrm{e}$ is
the electron rest mass \citep{2015MNRAS.447...49S}. 
We take $T_\mathrm{e}= T_\mathrm{gas}$ for simplicity.

We solve these equations in polar coordinate $(t,r,\theta,\phi)$ with
Kerr-Schild metric
by assuming axisymmetry with respect to the rotation axis,
$\theta=0$ and $\pi$.
The computational domain consists of
$r=[r_\mathrm{in},245r_\mathrm{g}]$, $\theta=[0,\pi]$.
Here we set the inner radius $r_\mathrm{in}$
to be $10\ \mathrm{km}$ for the NS
and $0.98 r_\mathrm{H}$ for the BH, where
$r_{\mathrm{H}}=M+(M^2+a^2)^{1/2}$ is a horizon radius
with $a$ being the spin parameter.
We take $a=0$ in this paper.
Numerical grid points are 
$(N_r, N_\theta, N_\phi)=(264, 264, 1)$. 
A radial grid size exponentially
increases with radius, and a polar grid is given by $\theta=\pi x_2 +
(1-h)\sin(2\pi x_2)/2$, where $h=0.5$ and $x_2$ spans uniformly from $0$
and $1$. We adopted outgoing boundary at outer radius, and reflective
boundary is adopted at $\theta=0$ and $\pi$. At the inner boundary
$r=r_\mathrm{in}$, a mirror symmetric boundary condition is employed for
the case of the NS,
while an outgoing boundary condition is used for the
the case of the BH.
That is, the matter as well as the energy is not swallowed
by the NS.

We start simulations from an equilibrium torus given by 
\cite{1976ApJ...207..962F}, but the gas pressure in this solution is 
replaced by a gas + radiation pressure by assuming a local
thermodynamic equilibrium. The inner edge of initial torus is situated at
$r=20r_\mathrm{g}$, while its pressure maximum is situated at
$33r_\mathrm{g}$.
Weak poloidal magnetic fields are initially embedded in the torus.
The magnetic flux vector $A_\phi$ is proportional to $\rho$,
and a ratio of maximum $p_\mathrm{mag}$ and
$p_\mathrm{gas}+p_\mathrm{rad}$ is set to be 100. 
Outside the torus, the gas is not magnetized and
the density and the pressure are given by 
$\rho=10^{-4}\rho_0 (r/r_\mathrm{g})^{-1.5}$
and $p_\mathrm{gas}=10^{-6} \rho_0 (r/r_\mathrm{g})^{-2.5}$,
where $\rho_0$ is the maximum mass density inside the torus.  
We also set $p_\mathrm{rad}=10^{-10} \rho_0$ 
and $u_\mathrm{rad}^\mu=(1,0,0,0)$ outside the torus.
\begin{figure*}
 \begin{center}
 \includegraphics[width=12cm]{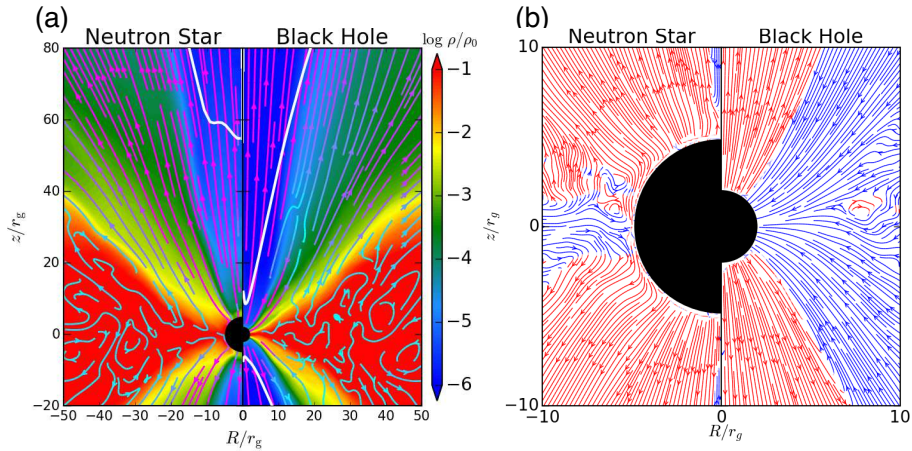}
  \caption{(a):Global structure of accretion disks and outflows for neutron
  star (left) and black hole (right) case. Color shows mass density,
  vectors shows stream lines, and white curves show photosphere. 
  (b): enlarged view of stream lines around the neutron star (left) and black hole
  (right). Red (blue) lines indicate that the radial velocity is in positive
  (negative) direction.} 
\label{fig1}
 \end{center}
\end{figure*}

In this paper, we take $\rho_0 = 0.1\ \mathrm{g\ cm^{-3}}$ for the
NS. On the other hand, the relatively small maximum
mass density is employed for the BH
($\rho_0=1.4\times 10^{-2}\ \mathrm{g\ cm^{-3}}$).
By such adjustment,
we can compare the models of NS and BH
under the almost equal condition,
since the mass of the NS 
is about one order of magnitude smaller than 
that of the BH.
In present work, we ignore the rotation of a central object ($a=0$). 
We also consider an unmagnetized NS. Thus we can directly 
study effects of physical boundary at a surface of central objects by 
comparing results between the BH and NS.

\section{Results}
\subsection{Overview of the two cases}\label{f1}
In the following, we show time averaged data between
$t=3,000t_\mathrm{g}-5,000t_\mathrm{g}$ at which the mass accretion
continuously occurs onto a central star.
We first give in Figure 1 global supercritical accretion flow patterns, comparing
the two cases of NS accretion (left) and BH accretion (right).
The color contours in figure 1-(a) represent gas density distribution
with the same color scales (but note that the density normalizations
$\rho_0$ is by a factor of $\sim$ 7 greater in the left panel), 
and arrows show fluid stream lines.
White lines indicate photosphere measured from outer boundary at
$r=245r_\mathrm{g}$ along fixed $\theta$.
The size of the NS (=10 km) corresponds to 4.8 $r_{\rm g}$ for a mass of
1.4 $M_\odot$.
Figure 1-(b) shows stream lines around the NS (left) and BH (right). Red and blue
lines indicate that the radial velocity is in positive and negative
direction, respectively. 

The flow patterns displayed in these figures are distinct in many respects.
First of all, the flow lines are roughly conical (i.e., the line
directions are more or less radial) in the innermost region (at $r
\lesssim 15 r_{\rm g}$) in the BH case (see the right panel),
while they are chaotic, especially in the innermost region in the NS
case (the left panel).
Second, the high density regions (indicated by the red color) is thinly collimated near the BH 
and thus has a conical structure in the BH accretion, 
while it is rather broadened and covers the large surface area of the NS.
Third, we see more significant outflow motion in the NS case.
In particular, the strong outflow is ejected even below the photosphere
(indicated by the thick white line).
The outflow has a large opening angle from $\simeq 60^\circ$
and its four velocity in orthonormal frame is $0.2$ around
$r=60_\mathrm{g}$ and $\theta = 60^\circ$, while it is only $0.005$ for
BH case. The mass flux is order of magnitude larger for NS than that of BH.
As these consequences, some of the inwardly flowing material in the NS
accretion flow 
does not reach the NS surface but is reflected and turns its direction
to outward.
No such reflection motion is significant in the BH
accretion flow (see figure 1-(b)).
These differences should be understood in terms of the different
mechanisms of absorbing radiation effects.

Figure 2 shows radial profiles of mass inflow rate $\dot
M_\mathrm{in}$ (red), outflow rate $\dot M_\mathrm{out}$ (blue), 
and net inflow rate $\dot M_\mathrm{net} = \dot M_\mathrm{in}-\dot
M_\mathrm{out}$ (black), for neutron star (solid) and black hole
(dashed).
For the NS, the mass inflow rate is about $\dot
M_\mathrm{in}\simeq 300 L_\mathrm{Edd}$ around $10r_\mathrm{g}$. 
It steeply decreases with a decrease in radius near the NS surface at $r=4.8r_\mathrm{g}$
since we employ reflection boundary condition. 
Also the mass outflow rate has a similar trend with that of the inflow rate, but it
is slightly smaller than the inflow rate. This indicates that
substantial mass is blown away from the disk. 
We note that the mass supply (inflow) rate around $r=20r_\mathrm{g}$ is about
$10^{3}L_\mathrm{Edd}$ in both case, since we start from the similar initial
torus. Even though that, the mass outflow rate is much higher for the NS than
that for BH. Thus, it indicates that the NS can drive more massive
outflows than the BH. We also note that the net inflow rate is
approximately constant inside $r\gtrsim 15r_\mathrm{g}$ for BH case. 
Thus, the inflow-outflow equilibrium is realized inside this radius.
For the NS case, the net inflow rate is not constant but it
slightly increases with increasing radius, even though the computational time is
the same ($t=3,000-5,000r_\mathrm{g}$) in both simulations. 
This would be due to the mass accumulations on the NS as shown above
(see also figure 1). 
To summarize, a fraction of about a few tens of percent of the input
mass can accrete onto a BH, whereas only ten percent of less of the
input mass can accrete onto a NS. The other fraction of mass is lost as
outflow.

\subsection{Various energy density distributions}\label{f2-3}
Next, we consider energy composition in the accretion disks with
different central objects. The kinetic, gas, magnetic and radiation
energy densities are expressed as
\begin{eqnarray}
 E_\mathrm{kin} &=& \rho (\gamma - 1)\gamma,\\
 E_\mathrm{gas} &=& (e + p_\mathrm{gas})\gamma^2 - p_\mathrm{gas}\\
 E_\mathrm{mag} &=& b^2 \gamma^2 - (n_\alpha b^\alpha)^2 \\
 E_\mathrm{rad} &=& n_\alpha n_\beta R^{\alpha \beta}
\end{eqnarray}
where $n_\alpha = (-\alpha,0)$ is the normal observer's four velocity,
$\alpha=(-g^{00})^{-1/2}$ is the lapse function, and 
$\gamma = - n_\alpha u^\alpha$ is the Lorentz factor.
The energy density is normalized by $\rho_0$.

Left three panels in figure 3 show spatial distributions of $E_\mathrm{kin}, E_\mathrm{mag}$ and
$E_\mathrm{rad}$. 
Again, the conical flow structure around the BH is clearly shown in the lower panels of figure 3
except for the magnetic energy distribution that shows a more spherically symmetric shape
(see the second panel from the left).
By contrast, the NS accretion case displayed in the upper panels show somewhat distinct pattern.
The upper, third panel from left, for example, show that 
the large $E_{\rm rad}$ region is found more widely around the NS than that around the BH.
This indicates intense radiation emitted from the NS surface and from the innermost flow region.
Kinetic energy distribution displayed in the upper left panel shows a similar structure,
implying launch of outflow occurring widely from the surface of the accretion flow.
Such enhanced energy regions around the central object are not found in the lower panels,
since excess energy can be absorbed by the BH.

Right panel in figure 3 shows comoving frame radiation energy density
distributions
normalized by $L_\mathrm{Edd}/(4\pi r^2 c)$, where we recover the light
speed $c$ for the sake of clarity.
We found that this quantity largely exceeds unity, typically $\sim 10^3$
or even greater, in the entire inflow region.
This is true in both of NS and BH cases, though the photon accumulation region
is much wider in the former. 
This fact indicates that there exists a region full of radiation energy
and that its radiation energy density is so high that it would be able to blow away
the large amount of gas by counteracting the gravity force. 
Nevertheless, we find that the inflow region stably persists around the compact objects. 
This is because the inflow exists deeply inside the photosphere (see Figure 1) so that 
the radiation flux can be much attenuated to become $F_{\rm rad} \ll E_{\rm rad}c$.
As a result, the gas is never prevented from accretion
\citep{2007ApJ...670.1283O}. 
This issue will be discussed later again.
\begin{figure}
 \begin{center}
 \includegraphics[width=9cm]{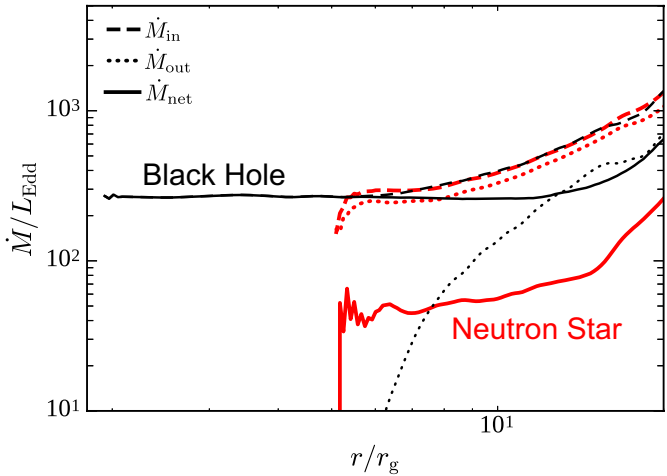}
  \caption{Radial profile of mass inflow rate (dashed), outflow rate
  (dotted), and net inflow rate (solid) for neutron star (red) and black
  hole (black).} 
\label{fig2}
 \end{center}
\end{figure}

Figure 4 shows the density weighted, angle-averaged energy densities in various forms along $r$.
We take an average of a physical quantity, $f$, over the entire solid angle ($\Omega$) according to
\begin{equation}
 <f> = \frac{\int d \Omega\ f \rho \sqrt{-g}}{\int d \Omega\ \rho \sqrt{-g}},
\end{equation}
where $g=\det\ g_{\mu\nu}$.

Comparing these panels, we understand that the kinetic energy $E_{\rm kin}$ 
dominates over all other energy forms inside the accretion disks in both cases.
While the radiation energy density $E_{\rm rad}$, the second largest one,
increases with decreasing radius in both cases, there exists an interesting distinction between the two:
the ratio of $E_{\rm rad}/E_{\rm kin}$ increases with a decreasing radius near the central object
in the NS accretion, while the opposite is the case in the BH accretion.
In the proximity of the NS, especially, the radiation energy density is comparable to 
the kinetic energy density (see also fig. 3). 
(Note that the kinetic energy is due mostly to the rotation, not to the
accretion.)
These facts indicate that the radiation pressure force makes a significant contribution in force balance near the NS
(this point will be discussed in the next subsection).
Around the BH, in contrast, the ratio of $E_{\rm rad}/E_{\rm kin}$ stays nearly constant
on the order of $\sim 10$ \% but rather decreases in the innermost part. 
This is the direct consequence of photons being swallowed by the BH.
We should note, however, that the difference between $E_\mathrm{kin}$ and $E_\mathrm{rad}$
may depend on the mass accretion rate.

The magnetic energy is unimportant in both cases;
the ratio of $E_{\rm mag}/E_{\rm kin}$ is always around a few \%.
Likewise, the gas energy $E_{\rm gas}$ is everywhere negligible because the gas temperature
is low enough.
An interesting distinction between the BH and NS cases is found regarding the magnetic energy distribution;
that is, it is nearly isotropic in the BH accretion while it is concentrated
on the polar and equatorial regions in the NS accretion (see Fig. 3). 
In our simulations, we start from the poloidal magnetic field. 
The magnetic flux is swept according to the gas accretion and it is accumulated near the central object. 
Since we assume ideal MHD and axisymmetry, the magnetic field is dissipated by a small
numerical resistivity and most of the flux remains around the pole.
\begin{figure}
 \begin{center}
 \includegraphics[width=8cm]{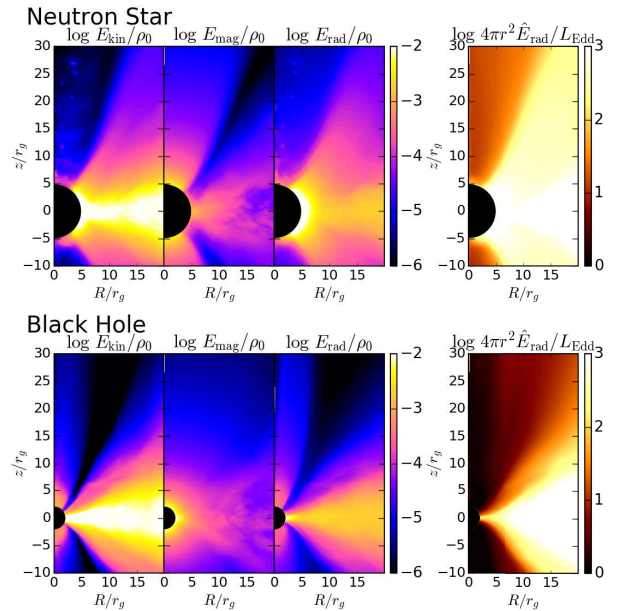}
  \caption{First three panels from the left show kinetic, magnetic, and
  radiation energy density. The right panel shows the comoving frame
  radiation energy density normalized by $L_\mathrm{Edd}/(4\pi r^2 c)$. Top and bottom panels
  correspond to the case for neutron star and black hole, respectively. } 
\label{fig3}
 \end{center}
\end{figure}

\begin{figure}
 \begin{center}
 \includegraphics[width=8cm]{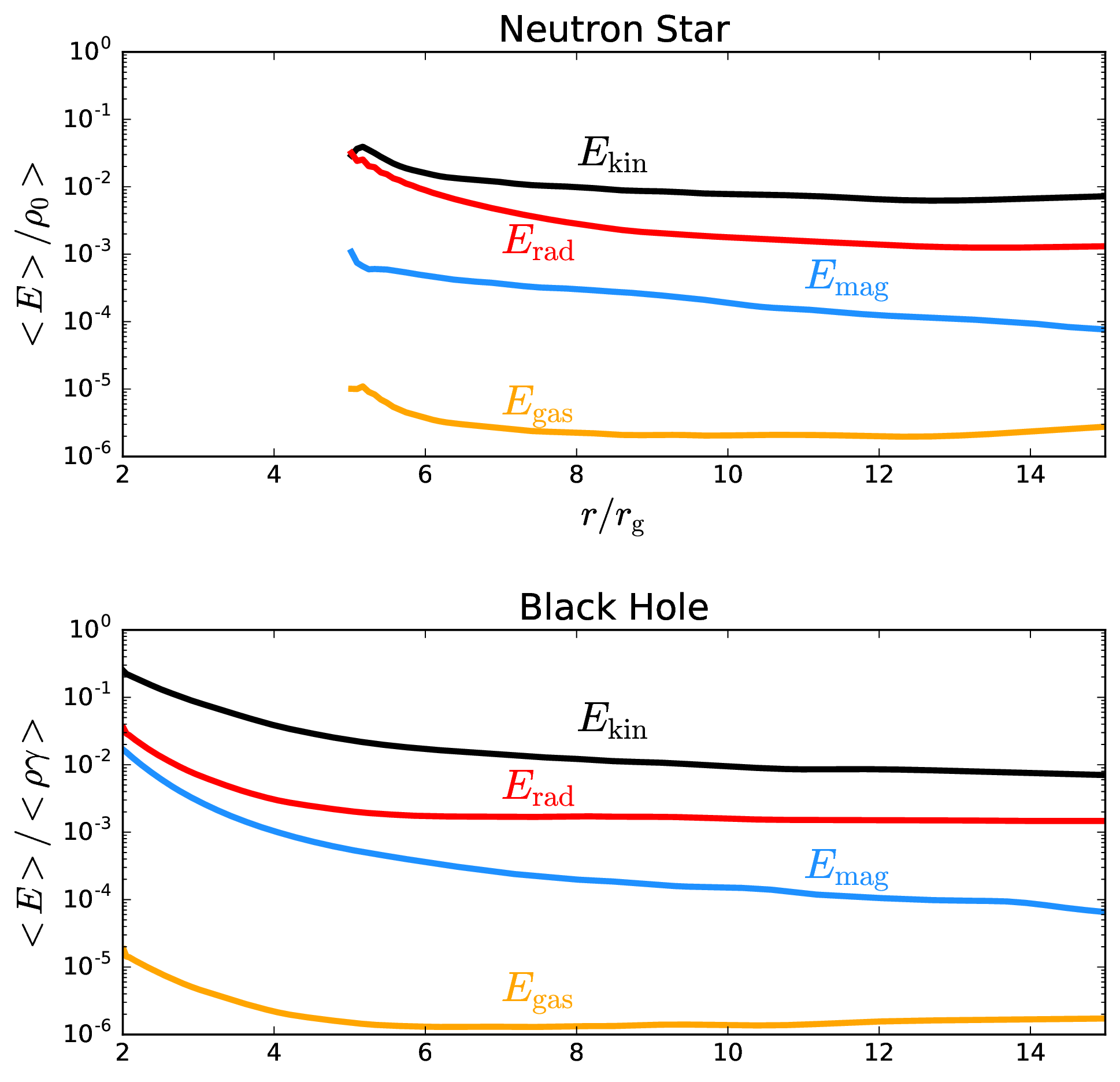}
  \caption{Density weighted kinetic (black), radiation (black), magnetic
  (blue) and gas energy density (orange). Top panel shows result for
  neutron star, and bottom does for black hole. } 
\label{fig4}
 \end{center}
\end{figure}

\subsection{Force balance on the equatorial plane}\label{f4}
Next we show the radial profile of forces acting on the fluid elements.
We consider a steady state equation of motion
\begin{eqnarray}
  f^\mathrm{adv}_r+
   f^\mathrm{grav}_r + f^\mathrm{cent}_r + f^\mathrm{rad}_r
  +f^\mathrm{gas}_r + f^\mathrm{mag}_r + f^\mathrm{cor}_r=0,\label{eq:fdcomp}
\end{eqnarray}
Here $f^\mathrm{adv}_r, f^\mathrm{grav}_r, f^\mathrm{cent}_r, f^\mathrm{rad}_r,
f^\mathrm{gas}_r, f^{\mathrm{mag}}_r, f^\mathrm{cor}_r$ are 
defined according to \cite{2015arXiv150906644M} as,
\begin{eqnarray}
 f^\mathrm{adv}_r &=&-u^j \partial_j u_r,\label{eq:fadv}\\
 f^\mathrm{grav}_r &=& \frac{T^t_t}{w}\Gamma^t_{rt},\label{eq:fgrav}\\
 f^\mathrm{cent}_r &=& \frac{T^\phi_\phi}{w}\Gamma^t_{rt},\label{eq:fcent}\\
 f^\mathrm{rad}_r &=& \frac{G_r}{w},\label{eq:rad}\\
 f^\mathrm{gas}_r &=& -\frac{\partial_r p_\mathrm{gas}}{w},\label{eq:fgas}\\
  f^\mathrm{mag}_r &=&-\frac{-\partial_r (b^2/2) + \partial_i (b^i b_r)}{w},\label{eq:fmag}\\
 f^\mathrm{cor}_r &=& f^\mathrm{metric}_r - f^{\mathrm{grav}}_r -
  f^{\mathrm{cent}}_r + f^\mathrm{ent}_r,\label{eq:fcor}
\end{eqnarray}
where
\begin{eqnarray}
 f^\mathrm{metric}_r &= & \frac{1}{w}T^\kappa_\lambda \Gamma^\lambda_{r\kappa}
  -\frac{T^i_r - \rho u^i u_r}{w}\frac{\partial_i
  \sqrt{-g}}{\sqrt{-g}},\\
 f^\mathrm{ent}_r &=& -\frac{u_r}{w}\partial_i\left[(w-\rho)u^i\right],
\end{eqnarray}
where $w = \rho + e + p_\mathrm{gas} + 2 p_\mathrm{mag}$ denotes the
relativistic enthalpy.
Here equations (\ref{eq:fadv}) - (\ref{eq:fmag}) correspond to
advection term, gravity
force, centrifugal force, radiation force, gas pressure gradient force
and Lorentz force. $f^\mathrm{cor}_r$ is the relativistic correction term.
\begin{figure}
 \begin{center}
 \includegraphics[width=8cm]{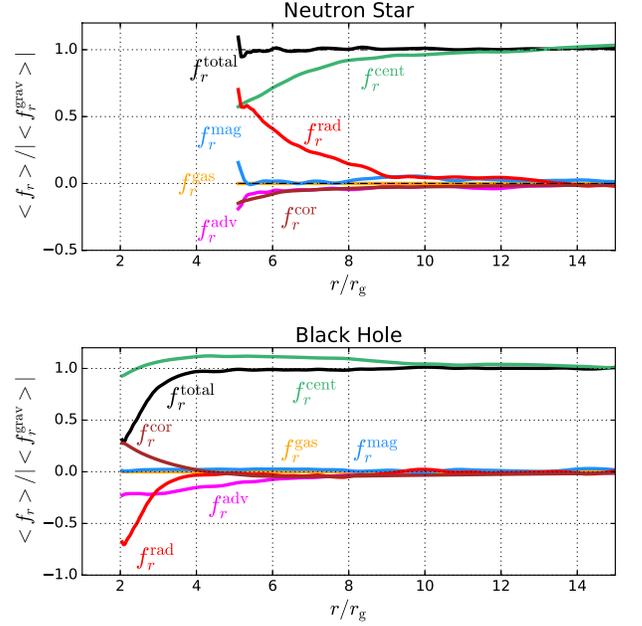}
  \caption{Density weighted radial force normalized by the gravity
  force.} 
\label{fig5}
 \end{center}
\end{figure}

Figure 5 shows various forms of density weighted, angle-averaged radial force along $r$ 
normalized by the gravity force. 
Here $f^\mathrm{tot}_r$ is the total force without gravity force, so that
steady accretion would be realized where $f^\mathrm{tot}_r/|f^\mathrm{grav}_r| = 1$.
Let us first examine the NS case displayed in the upper panel.
We immediately notice that the centrifugal force balances almost completely with the gravity force
at large radii far from the central object. Hence, the rotation profile is nearly Keplerian and
radiation force is negligible there. 
With a decrease in radius, however, the outward radiation force grows, since the NS surface
cannot swallow the radiation so that the radiation energy is accumulated there. 
The radiation energy density profile, hence, has a negative gradient along $r$, 
which gives rise to outward radiation pressure force. 
The centrifugal force decreases with a decreasing radius so that 
the radiation force and centrifugal force can be comparable close to the NS surface.
This occurs because the gravitational attraction force by the NS is weakened by the outward 
radiation-pressure force. As a result, the disk rotation becomes highly sub-Keplerian,
although the flow is still in a quasi-steady state.
The important fact is that radiation force does never exceed the gravitational force,
which makes it feasible to induce supercritical accretion flow.

It is then of great importance to pay attention to the behavior of the centrifugal force.
We find a clear tendency that it declines inward very close to the NS.
This is caused by the accumulation of low angular momentum above the NS surface
and never happens in the BH case, since matter should be immediately swallowed.
But the gradient of the radiation energy density is not large enough to totally compensate
the gravitational attraction force towards the NS.
Finally, the advection term is very small, compared with the gravity force, but it does not vanish.
That is, the matter is slowly accreting onto the NS surface with accretion velocity
being much less than the free-fall velocity. 
We may call this slowly accreting zone (at $r < 10 r_{\rm g}$) the settling region.
As a result, the supercritical accretion is feasible for the NS.

Next, let us examine the force balance in the BH accretion case in comparison with the NS case.
A big distinction is found in the behavior of the radiation force, 
which is negative in the BH case, while it was positive in the NS case.
This is because not only the gas but also the radiation energy is swallowed by the BH.
The negative radiation flux pushes the gas toward the BH.
This explains why supercritical accretion onto a BH is feasible 
\citep[see][but for the discussion based on the pseudo-Newtonian dynamics]{2007ApJ...670.1283O}.

Another distinction is that there is no force balance near the BH
in the sense that the total force does no longer balance with the gravity force near the BH.
This means, mass is continuously falls onto the BH with finite velocity.
Especially, the accretion motion is supersonic and is close to speed of
light in the BH vicinity.

We note that the centrifugal force exceeds the gravity force inside
$r<10r_\mathrm{g}$ for BH, but the total force balance holds if we
consider the relativistic correction factor $f_r^\mathrm{cor}$, i.e., a
quasi-steady state does actually realizes. There is an issue how we
decompose each force term in equation (\ref{eq:fdcomp}). 
The centrifugal force $f^\mathrm{cent}_r$ approaches to non-relativistic
one far from the black hole, but this force does not balance with gravity
force everywhere. It deviates from the gravity force close to the
central object. The relativistic correction term $f_r^\mathrm{cor}$ is
important in this region. 
For example, the innermost stable circular orbit is never obtained without
$f_r^\mathrm{cor}$. 
The gravity force almost balances with the centrifugal force and
correction force in this region, but the advection and radiation forces
are also important and thus, the total force balances with the gravity force.

\section{Discussion}
In the present paper we have carefully examine the gas dynamics of supercritical flow
around the NS, in comparison with that around the BH, through the GR-RMHD simulations.
Supercritical accretion is feasible in both of NS and BH cases but for distinct reasons.
While it is photon trapping that works in the BH case, the removal of mass and energy 
in the form of intense outflow is a key to realizing supercritical accretion onto NS.
The flow dynamics is also distinct:
sub-sonic, settling flow occurs around the NS surface, whereas matter nearly free falls onto the BH.
In the following, we will discuss some related issues more or less related to supercritical NS accretion.

\subsection{Outflow from inside the spherization radius}
It is widely known that SS73 have proposed the standard disk model,
but in the same paper they also made pioneering discussion regarding the gas dynamics of the
supercritical accretion flow onto the BH. 
In their section IV, SS73 introduced the notion of the spherization radius, 
inside which gas flows towards the central BH in a spherically symmetric fashion.
They also pointed out that outflow emerges from inside this radius.
They evaluated the spherization radius to be on the order of
$r_{\rm sph} \sim 10 ({\dot M}c^2/L_{\rm Edd}) r_{\rm g}$,
corresponding to the trapping radius, inside which photon trapping is significant (see also Begelman 1982).
In the present case we estimate $r_{\rm sph} \sim 10^3 r_{\rm g}$ 
(for ${\dot M}c^2/L_{\rm Edd}\gtrsim 300$, see Figure 2)
thus being far outside the picture box of figure 1.

The right panel of figure 1 clearly shows 
that the inflow and outflow streamlines are separated all the way down to the BH event horizon.
In other words, there are no stream lines which turns its direction from inward to outward.
By contrast, the left panel of figure 1 shows somewhat similar streamlines as those illustrated in Fig. 8 of SS73;
that is, some streamlines change their directions from inward to outward. 
Rather, we see that the change of the direction occurs even in the very vicinity of the NS surface.
In fact, the inflow and outflow rates nearly coincide in the innermost region
(inside $\sim 10 r_{\rm g}$, see, figure 2) so that the net accretion
rate is kept around the critical rate.
This is exactly a situation as that postulated by SS73.
\begin{figure}
 \begin{center}
 \includegraphics[width=8cm]{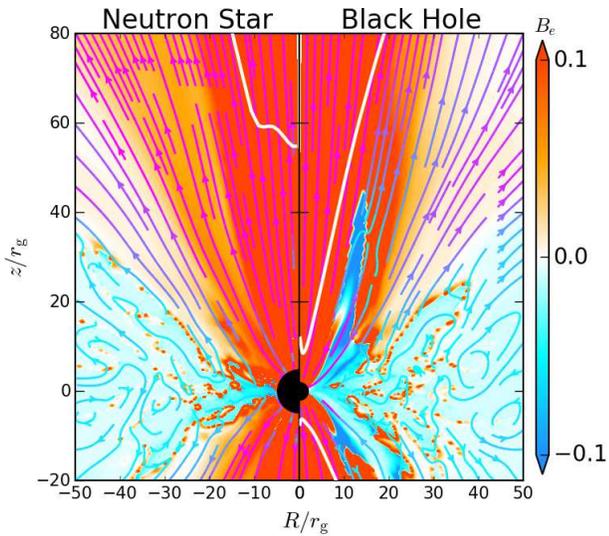}
  \caption{Same with figure 1, but color shows Bernoulli parameter. } 
\label{fig6}
 \end{center}
\end{figure}

\subsection{Bernoulli parameter}\label{f5}
To visualize the relative importance of the outflow in the NS accretion,
we calculate the local Bernouilli parameter according to Sadowski \& Narayan (2015);
\begin{equation}
   {\rm B_e} \equiv -\frac{T_t^r + R_t ^ r + \rho u^r}{\rho u^r},
\end{equation}
where $T_t^r$ and $R_t^r$ are the $t-r$ components of the MHD and
radiation energy-momentum tensors
(representing the energy flux of MHD and radiation processes), respectively, 
and $\rho u^r$ stands for the rest-mass energy flux. 

The results are shown in figure 6 for the NS and BH cases in the left and right panels, respectively. 
The locations of the photospheres are also indicated by the thick white lines there.
It is obvious that the blue regions, in which ${\rm Be} < 0$, are wider in the BH case.
Especially, we find that Bernouilli parameter is negative mostly below the photosphere close to the BH,
while it is positive in the NS case (except near the equatorial plane).

\subsection{Radiation cushion}
A next question which we wish to address is if there exists a settling regime covering the NS surface.
The accretion column created on the magnetized NS surface is composed of the upper free-fall region 
and the lower settling region (e.g. Basko \& Sunyaev 1976, Kawashima et al. 2016).
In the latter, accretion velocity is much reduced by the decelerating force asserted by radiation cushion.

The direct consequence of the existence of the settling region is that
the matter density is $\rho \propto r^{-3}$, radiation pressure is $P_{\rm rad} \propto r^{-4}$,
and radiation temperature is $T_{\rm rad} \propto r^{-1}$.
These relations are derived from the hydrostatic balance in the radiation-pressure dominated atmosphere,
which leads
\begin{equation}
    \frac{GM\rho}{r^2} = - \frac{dP_{\rm rad}}{dr}.
     \label{eq:fbalance}
\end{equation}
Here, we assume that accretion motion is very slow (accretion velocity is much less than free-fall velocity).
Let us further assume little entropy production is significant during the accretion.
Then, the adiabatic relation holds between $P_{\rm rad}$ and matter density $\rho$; that is,
$P_{\rm rad}\propto \rho^{4/3}$. We then find
$dP_{\rm rad}/P_{\rm rad}^{3/4} \propto dr/r^2$, which reads $P_{\rm rad} \propto r^{-4}$
and $\rho \propto P_{\rm rad}^{3/4}\propto r^{-3}$. 

To see if such dependences appear in the simulation data of the NS case,
we plot matter density and $(T_\mathrm{rad})^3$ as functions of radii in figure 7.
We find that radiation entropy crudely obeys the expected relationship;
$T^{3}_{\rm rad} \propto r^{-3}$ in the innermost region,
$r < 10 r_{\rm g}$,
although the density profile is steeper than $r^{-3}$.
These results indicate an almost adiabatic settling region is
formed close to the NS.
The mass density and radiation entropy on the surface of NS
increase with time due to the accumulation. Nevertheless, their radial
profiles do not change.
This indicates the force balance given in equation (\ref{eq:fbalance})
holds during simulation interval. Thus, we can expect that supercritical
accretion onto the NS continues in accompany with forming settling
region, until the gas in the disk is exhausted and mass accretion rate decreases.

\begin{figure}
 \begin{center}
 \includegraphics[width=8cm]{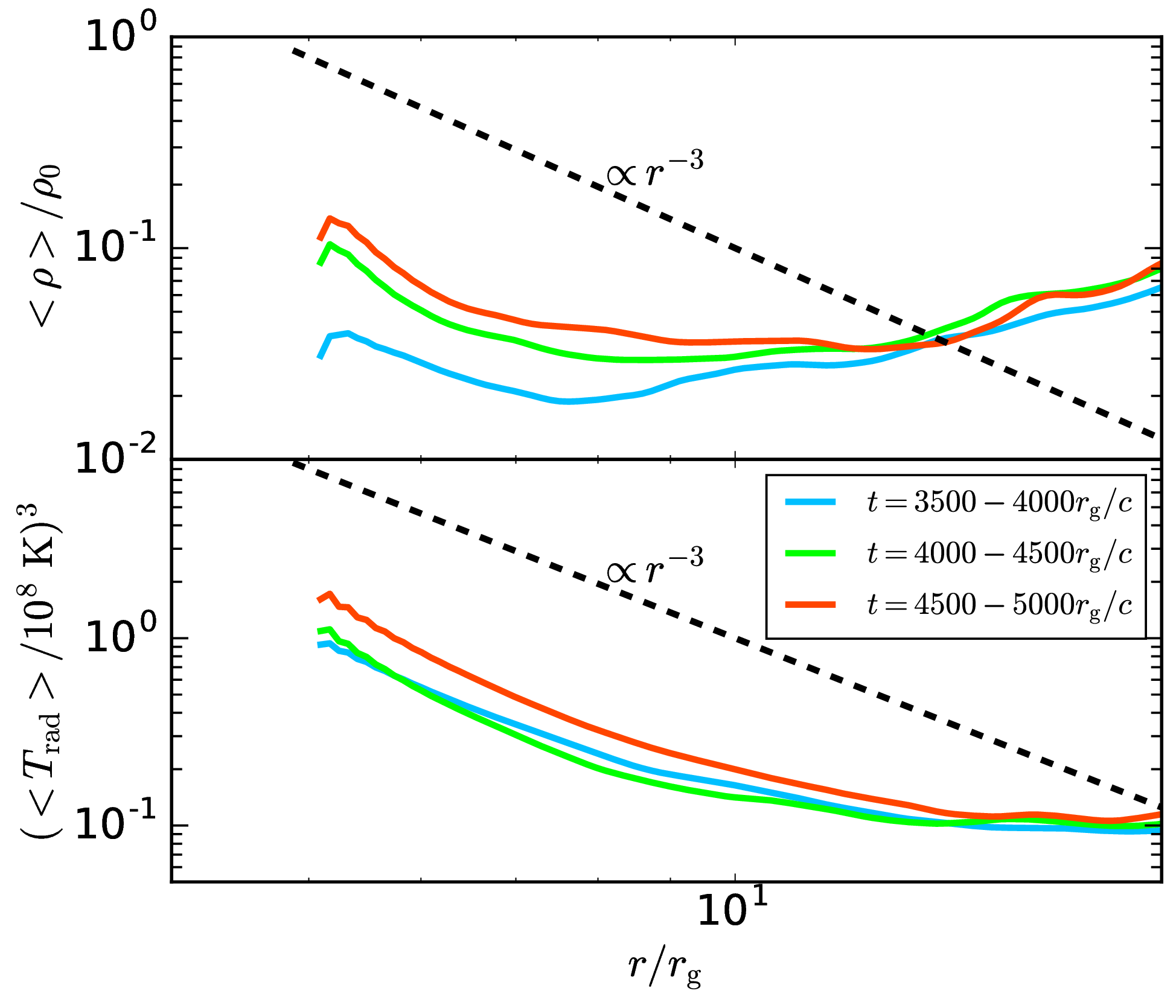}
  \caption{Density weighted $\rho$ and $T_\mathrm{rad}^3$ profiles with
  different time intervals. }
\label{fig7}
 \end{center}
\end{figure}

  \subsection{Validity of our numerical model}
We simply compute opacities assuming fully ionized hydrogen gas.
The free-free opacity is, however, much larger by assuming the solar
opacity. We expect results would not be affected so much by the
metallicity since the
local thermodynamic equilibrium ($T_\mathrm{gas}=T_\mathrm{rad}$) is
attained mainly due to the Comptonization whose cooling timescale is much
shorter for the supercritical accretion disks. For the scattering opacity, it
decreases about 15\% assuming the solar abundance. 
The reduction of opacity might reduce the outflow power. But the outflow
velocity is determined by the balance between the radiation force
($\propto \kappa_\mathrm{sca} F_\mathrm{rad})$ and
its drag force \cite[$\propto \kappa_\mathrm{sca} E_\mathrm{rad}$,
see][]{2015PASJ...67...60T}. The resulting terminal
velocity would not be affected by the opacity. Also
\cite{2005ApJ...628..368O} shows that the luminosity weakly depends
or is almost independent from the metallicity. Thus, our conclusion would
hold even if we adopted the solar metallicity.

Another concern in our numerical model is the boundary condition
on the neutron star. We simply applied a mirror boundary condition where
the gas never
flows across the boundary. This boundary condition might be plausible to
mimic the neutron star's solid surface,
while other boundary conditions have been adopted in the past study;
e.g., free boundary condition \citep{2012MNRAS.421...63R} or the
accretion-energy-injection boundary condition \citep{2007PASJ...59.1033O}.
Also the boundary condition adopted in our simulation does not take into
account the interaction between the gas and neutron star.
The magnetic activity in this boundary layer can transport the angular momentum 
\citep{2002MNRAS.330..895A}. The dissipation of rotation energy of the
disk would increase the radiation energy close to the neutron star.
Although recent high resolution MHD simulations show that the stresses
worked in the boundary layer oscillate around zero
\citep{2012ApJ...751...48P,2017arXiv170901197B},
it is under debate what boundary condition is appropriate to describe
the neutron star surface.
We have to perform comprehensive study around the neutron star surface
with different boundary condition models to investigate the plausible
boundary conditions. We remain this problem as a important future work.

\section{Conclusions}
We performed 2-dimensional axisymmetric GR-RMHD simulation of
supercritical accretion onto a non-rotating unmagnetized neutron star,
and comparing results with non-rotating black hole.  
Our findings can be summarized as follows:
\begin{itemize}
 \item In contrast with the black hole case, a significant fraction of
       mass is blown away by the radiation pressure driven outflow and
       thus the net mass inflow rate reduces for the neutron star
       case. Also the anisotropic radiation arising from the anisotropic
       density distribution helps photons escape from the disk.
\item Inside the accretion disks, the radiation flux is largely
      attenuated so that the radiation force balances with the sum of 
      centrifugal and gravity forces. Due to the large optical depth in
      the supercritical disks, the radiation energy density much exceeds
      that expected from the Eddington luminosity, $E_\mathrm{rad}\simeq
      F_\mathrm{rad} c/\tau > 100 L_\mathrm{Edd}/(4\pi r^2 c)$. 	     
 \item We found that the gas and radiation is accumulated on the neutron
       star surface. The settling region, where accretion motion is
       significantly decelerated by radiation cushion is formed. The
       radiation cushion would be approximately adiabatic, i.e., the
       radiation energy roughly follows $\hat E\propto r^{-4}$ and the
       gas and radiation temperature obeys $\propto r^{-1}$. Such a
       radiation cushion never appears around the black hole so that
       matter can be directly swallowed by the black hole.
       Also, these mass density and radiation energy density profiles
       follow radiation pressure supported hydrostatic balance.
\end{itemize}
These facts make supercritical accretion feasible for the neutron star. 
\acknowledgments
Numerical computations were carried out on Cray XC30 at the Center for
Computational Astrophysics of National Astronomical Observatory of
Japan, and on K computer at AICS.
This work is supported in part by JSPS Grant-in-Aid for 
Young Scientists (17K14260 H.R.T.) and for Scientific Research (C)
(17K05383 S. M., 15K05036 K.O.).
This research was also supported by MEXT as 'Priority Issue on Post-K
computer' (Elucidation of the Fundamental Laws and Evolution of the
Universe) and JICFuS.


\end{document}